\documentclass[9pt,conference]{IEEEtran}
\IEEEoverridecommandlockouts

\usepackage{cite}
\usepackage{amsmath,amssymb,amsfonts}
\usepackage{algorithmic}
\usepackage{graphicx}
\usepackage{textcomp}
\usepackage{xcolor}
\usepackage{booktabs}
\usepackage{multirow}
\usepackage{pifont}
\usepackage{tipa}
\usepackage{subfigure}
\def\BibTeX{{\rm B\kern-.05em{\sc i\kern-.025em b}\kern-.08em
    T\kern-.1667em\lower.7ex\hbox{E}\kern-.125emX}}
\newcommand{\cmark}{\ding{51}}%
\newcommand{\xmark}{\ding{55}}%
\begin{document}

\title{Weakly Supervised Phonological Features for Pathological Speech Analysis\\}


\author{
    \IEEEauthorblockN{
        Jenthe Thienpondt\textsuperscript{\textdagger}\thanks{\textsuperscript{\textdagger}Equal contribution.}\qquad
        Geoffroy Vanderreydt\textsuperscript{\textdagger}\qquad
        Abdessalem Hammami\qquad
        Kris Demuynck
    }
    \IEEEauthorblockA{
        \textit{IDLab, Department of Electronics and Information Systems, Ghent University - imec, Ghent, Belgium} \\
        jenthe.thienpondt@ugent.be
    }
}

\maketitle


\begin{abstract}
Paralinguistic properties of speech are essential in analyzing and choosing optimal treatment options for patients with speech disorders. However, automatic modeling of these characteristics is difficult due to the lack of labeled speech datasets describing paralinguistic properties, especially at the frame-level. In this paper, we propose a weakly supervised training method which exploits the known acoustic properties of phonemes by training an ASR model with an interpretable frame-level phonological feature bottleneck layer. Subsequently, we assess the viability of these phonological features in speech pathology analysis by developing corresponding models for intelligibility prediction and speech pathology classification. Models using our proposed phonological features perform similar to other state-of-the-art acoustic features on both tasks with a classification accuracy of 75\% and a 8.43 RMSE on speech intelligibility prediction. In contrast to others, our phonological features are text-independent and highly interpretable, providing potentially useful insights for speech therapists.
\end{abstract}


\section{Introduction}
Speech pathologies include a wide array of disorders that profoundly influence the ability to communicate. The originating causes are various and include surgical interventions in the vocal tract (e.g. laryngectomy, glossectomy), reduced motor speech capabilities (e.g. dysarthria) or as a developmental side-effect following a hearing disorder.

Analysis of pathological speech and the subsequent treatment options are traditionally determined by a subjective assessment of speech- and language-therapists~(SLTs). However, perceptual evaluation of speech by SLTs is error-prone and resource intensive, contributing towards the interest of automatic and objective speech evaluation methods~\cite{path_human_eval_int_non_reliable}. In this regard, methods relying on automatic speech recognition (ASR) models have proven effective, as objective speech intelligibility metrics based on ASR transcription accuracy have shown to correlate strongly with the perceptual evaluation of speech therapists~\cite{path_asr_2008, asr_int_overview, asr_int_clinical_1}. However, some major disadvantages are still present in ASR-based systems: the method requires a reference text and interpretability of the results is constrained to a linguistic analysis of the transcription. In addition, ASR-based metrics have only shown satisfactory results on the intelligibility regression task, covering only a small area of speech pathology analysis. Some promising results have been established by using latent representations of speech from deep neural networks (DNNs) for both pathology classification and intelligibility prediction~\cite{van_son_asr_int_cancer, path_xvec_int}. For example, speaker embeddings derived from the popular x-vector~\cite{x_vectors} or ECAPA-TDNN~\cite{ecapa_tdnn} models are commonly used as a text-independent alternatives for ASR-based metrics. However, models employing speaker embeddings have limited interpretability, greatly reducing their usefulness in a clinical context. 

To alleviate these issues, we propose a weakly supervised training method to extract frame-level phonological features (PLFs) from speech which combines the high predictive capacity of modern DNN-based features with high interpretability, essential for proper pathological speech analysis. Initially, the extraction of frame-level phonetic properties was introduced to improve the performance of downstream phoneme recognition models~\cite{plfs_asr_1994, jp_asr_plfs}. Some initial attempts have been made to use these phonetic properties for subsequent usage in pathological speech analysis~\cite{jp_plf_asr_dependent, jp_plf_asr_free} but their application was limited to intelligibility prediction and lacks a comparison with current state-of-the-art ASR- and DNN-based features. To address these shortcomings, we establish the feasibility of our proposed PLFs for both intelligibility prediction and speech pathology classification and provide a comparison with current state-of-the-art features commonly used in the literature. In addition, we compare the behavior of individual PLFs with current clinical research to asses their potential as an objective paralinguistic measurement.

The rest of the paper is organized as follows: section~\ref{s:weakly_supervised_plfs} describes our proposed weakly supervised PLF training setup and utterance-level feature extraction methods. Subsequently, section~\ref{s:experimental_setup} introduces our experimental setup for intelligibility prediction and pathology classification using the proposed PLFs. Section~\ref{s:results} continues with the corresponding results and analysis and is followed by the concluding remarks in section~\ref{s:conclusion}.

\section{Weakly Supervised Phonological Features}
\label{s:weakly_supervised_plfs}
\begin{figure*}[t]
\includegraphics[width=1.0\textwidth]{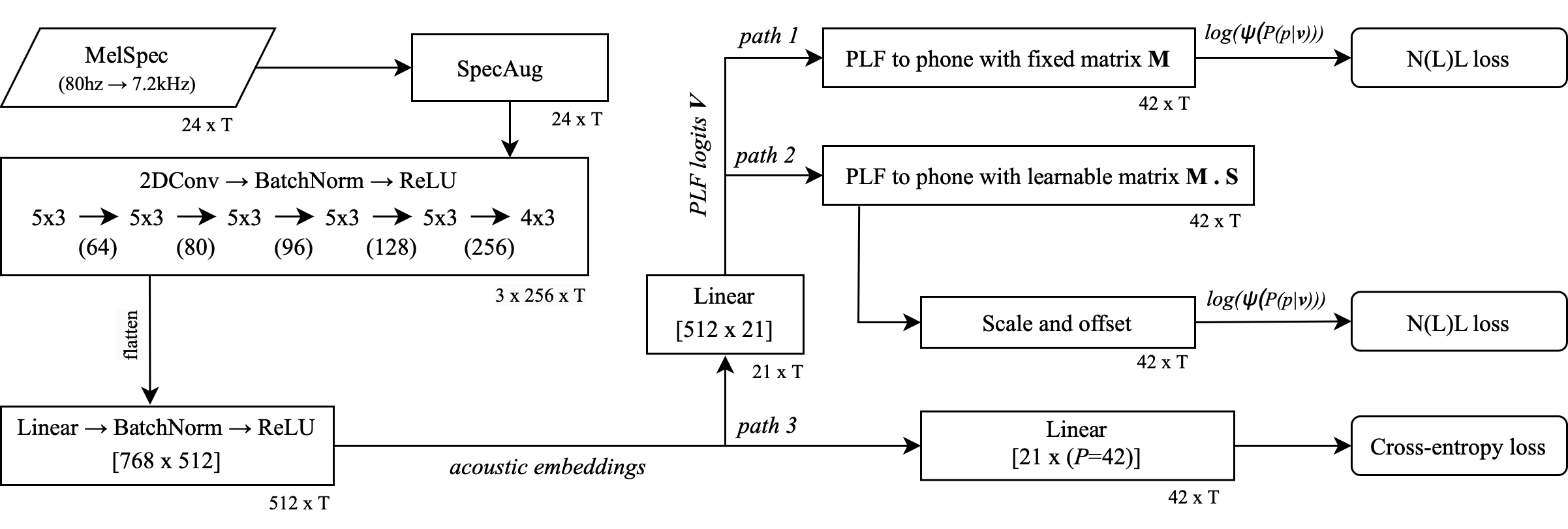}
\caption{Overview of training stage for our proposed phonological features. \textit{N(L)L} refers to negative (log-)likelihood, depending on the value of $E$.}
\label{fig:plf_training}
\end{figure*}

In this section, we will describe the two major stages in our proposed PLF framework. The first stage is training the frame-level PLFs with the guidance of the phonetic transcription of the training dataset and is depicted in Fig.~\ref{fig:plf_training}. The second stage converts the frame-level PLFs to an utterance-level representation for usage in downstream modeling tasks.

\subsection{PLF Training Setup}
\label{ss:plf_training_setup}
Our proposed\let\thefootnote\relax\footnotetext{Supported by Research Foundation Flanders (FWO) grant S004923N and EU Horizon 2020 programme TAPAS under Marie Curie grant 766287.} PLF training framework consists of a front-end feature extractor, which transforms the input audio to latent acoustic frame-level embeddings, followed by a PLF back-end, projecting the acoustic embeddings to $F$ phonological features which are subsequently used to classify $P$ phonemes. The input features for the front-end module consist of 24-dimensional Mel-spectrograms extracted from the input audio with a 32 ms window length and a 10 ms frame shift. Subsequently, SpecAugment~\cite{specaugment} is applied to prevent overfitting. The front-end feature extractor is composed of stacked 2D-convolutions followed by a fully-connected layer, transforming the frame-level features to 512-dimensional acoustic embeddings. More details of the front-end module are depicted in Fig.~\ref{fig:plf_training}.

\begin{figure}[t]
\begin{minipage}[b]{1.0\linewidth}
  \centering
  \centerline{\includegraphics[width=9cm]{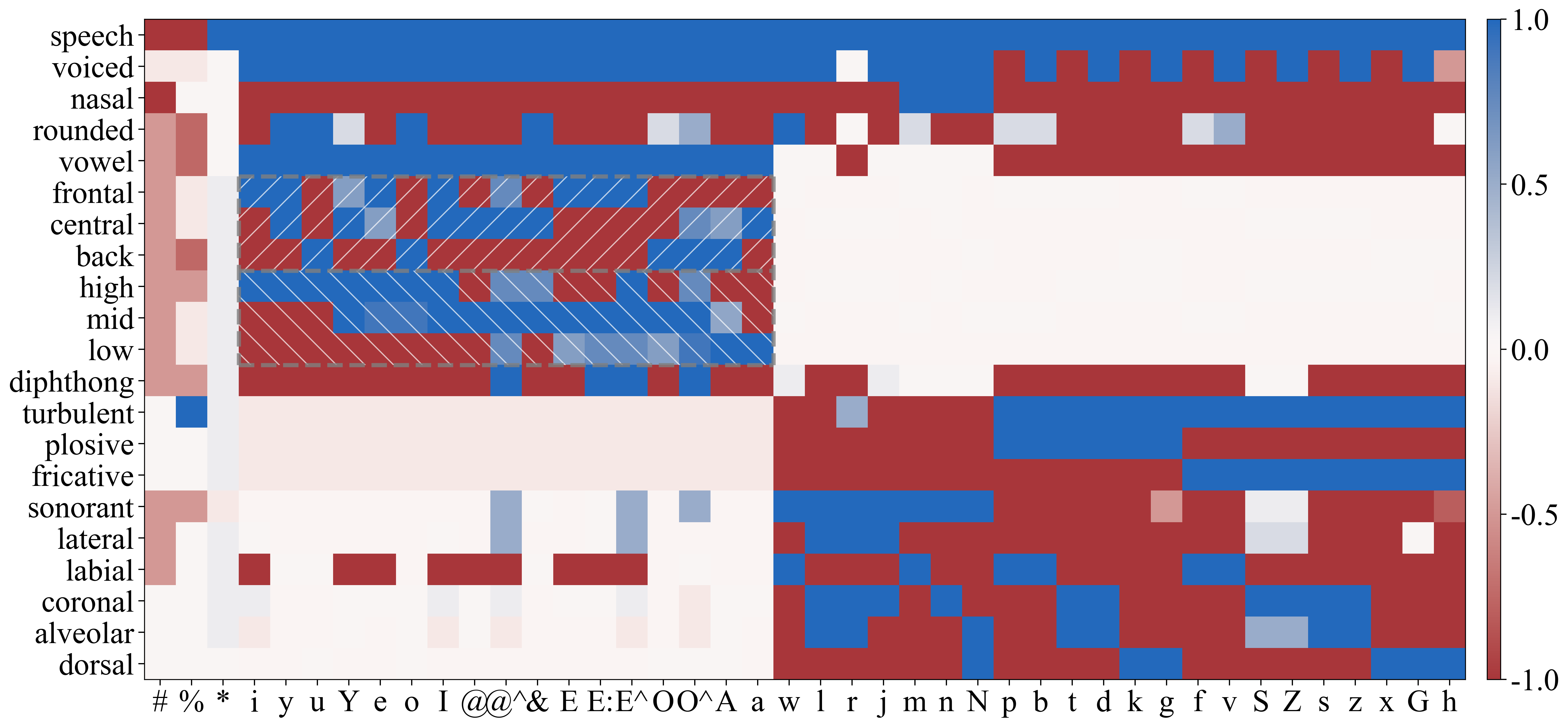}}
\end{minipage}
\caption{PLF to phoneme conversion matrix $\pmb{M}$. Striped regions indicate grouped attributes for the horizontal ($/$) and vertical ($\setminus$) vowel position PLFs.}
\label{fig:plf_to_phone_matrix}
\end{figure}

The training of the PLF-backend is guided by three paths with distinct losses to model robust phonological features. The first path projects each acoustic embedding to the PLF vector $\pmb{v}$ $ \in \mathbb{R}^{F \times 1} $, representing the logits of the phonological features. Subsequently, the PLF logits are converted to phone posterior probabilities by using the PLF-to-phone conversion matrix $ \pmb{M} \in \mathbb{R}^{P \times F} $ as depicted in Fig~\ref{fig:plf_to_phone_matrix}. The values in the conversion matrix represents the expected response of our 21 defined PLFs for each phone and can range from $-1$ (not active) to $1$ (active), with $0$ representing irrelevancy. Subsequently, we calculate the posterior probability $P(f \mid p)$ of a PLF $f$ having the expected value for each phoneme $p$ as follows:
\begin{equation}
\label{eq:plf_posterior}
  P(f \mid p) =
  \begin{cases}
    \sigma(v_f) & \text{if $M_{p, f} \geq 0$} \\
    \sigma(-v_f) & \text{if $M_{p, f} < 0$}
  \end{cases}
\end{equation}
All PLFs are updated independently during training except the PLFs related to horizontal (front, central or back) and vertical (high, mid or low) vowel position (indicated by the striped regions in Fig.~\ref{fig:plf_to_phone_matrix}). For these two grouped PLFs, we take a weighted sum of the posterior probabilities of their corresponding attributes as calculated in equation~\ref{eq:plf_posterior}, with the weight being determined by their matching $M_{p, f}$ value. This allows us to define a more granular vowel position needed to model all phonemes correctly. Additionally, it also enables the ability to model positional vowel transitions within a single phoneme, as is present in e.g. diphthongs. Subsequently, we estimate the posterior log probability $log(P(p \mid \pmb{v}))$ of phone $p$ as:
\begin{equation}
\label{eq:phone_prob}
    log(\psi(P(p \mid \pmb{v}))) = \sum_{f}^{F} \psi(log(P(f \mid p))) \cdot M_{p, f}
\end{equation}
The compression function $\psi(x)$ in equation~\ref{eq:phone_prob} allows us to interpolate between two viewpoints on how PLFs determine the posterior probability of a phoneme and is defined as:
\begin{equation}
\label{eq:exp_trick}
    \psi(x) = (\exp(\frac{x}{E}) - 1) \cdot E
\end{equation}
When $E$ is large, equation~\ref{eq:plf_posterior} will be reminiscent of a multiplicative model where each PLF must have the correct value for a corresponding phoneme. When $E$ is small, the expression turns towards an additive model, with a PLF having the correct value only contributing towards evidence of a specific phoneme while being more tolerant towards incorrect PLFs. The weight of grouped PLFs in equation~\ref{eq:phone_prob} is fixed to 1. Finally, the parameters are updated by minimizing the corresponding negative likelihood or log-likelihood, depending on the value of $E$.


The PLF-to-phone conversion matrix imposes a strict and manually determined mapping between PLFs and phonemes. Therefore, we allow in a second path small deviations from $\pmb{M}$ by employing a trainable scaling matrix $\pmb{S}~\in~\mathbb{R}^{P \times F} $ with $\forall s_{pf} > 0 $. Subsequently, the phone posteriors are calculated in equations~\ref{eq:plf_posterior} and~\ref{eq:phone_prob} using $\pmb{M}~\cdot~\pmb{S}$ as the conversion matrix. By only allowing positive scalars, the expected presence of a PLF for a certain phoneme according to $\pmb{M}$ is preserved while permitting the weight to change. Additionally, a trainable offset and scaling factor is employed on the final log phone posteriors to compensate for the lack of prior phoneme probabilities and an incorrect feature independency assumption, respectively. 

Finally, a third path simply calculates the cross-entropy from the predicted phone logits derived directly from the acoustic embedding. This is reminiscent of a traditional ASR training setup and mainly serves to ensure a robust acoustic embedding.


\subsection{PLF Feature Extraction}
\label{ss:plf_feature_extraction}
\begin{figure}[t]
    \centering
    \subfigure[Speakers pronouncing 'zoon']{%
        \includegraphics[width=0.48\textwidth]{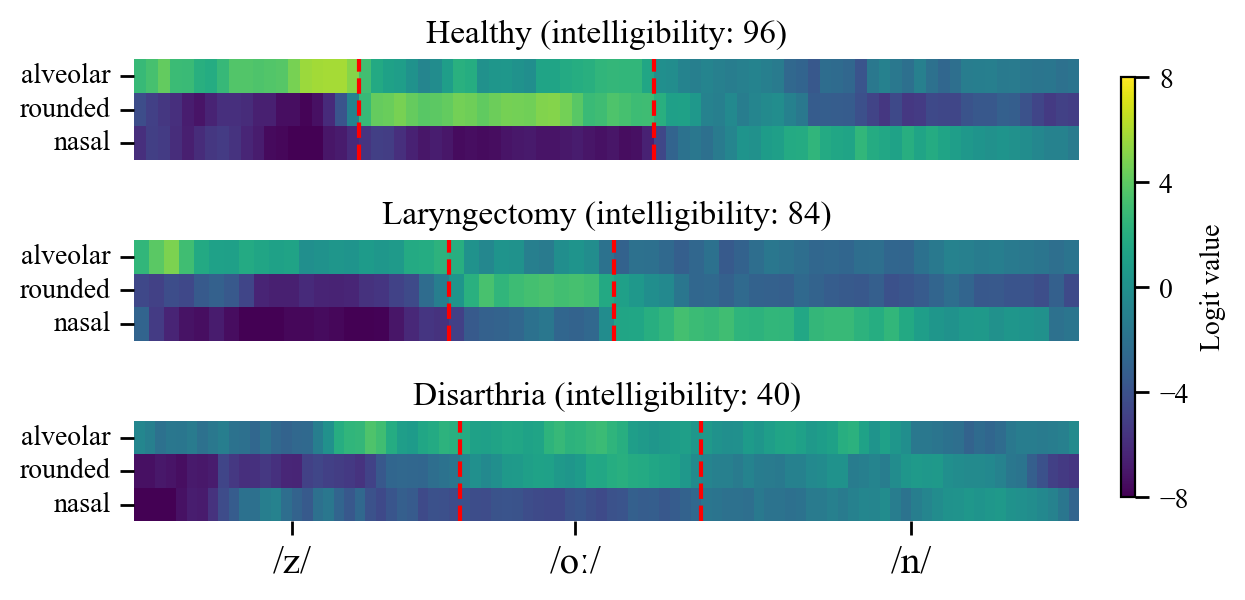}
        \label{fig:sample1}
    }
    \hfill
    \subfigure[Speakers pronouncing 'wook']{%
        \includegraphics[width=0.48\textwidth]{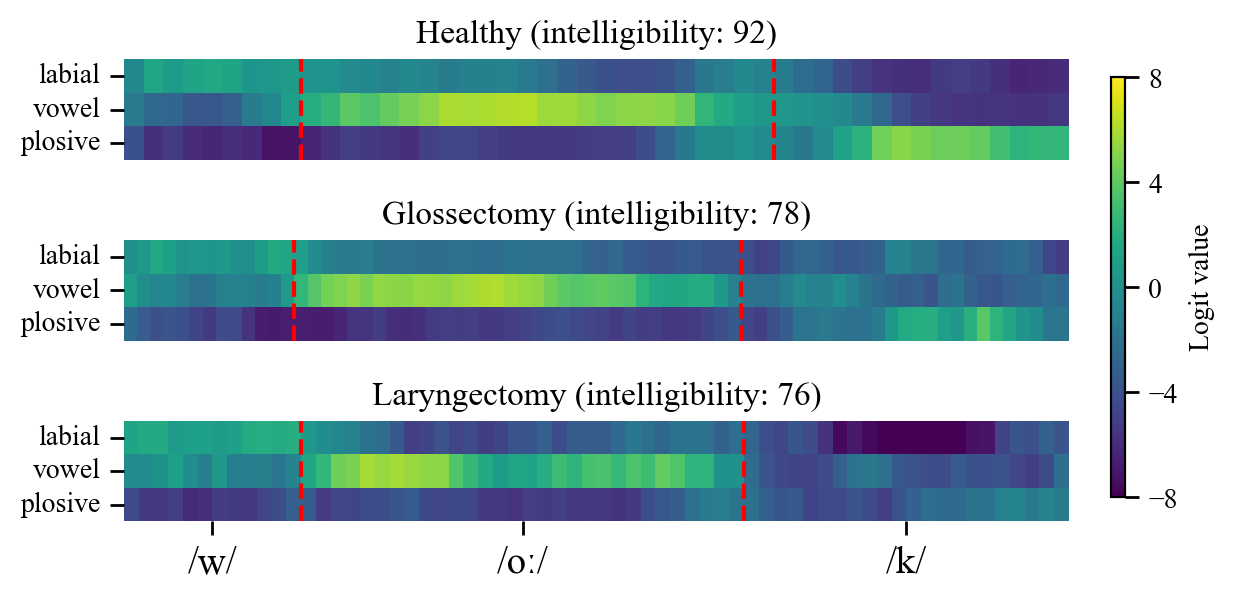}
        \label{fig:sample2}
    }
    \caption{Frame-level PLF logits of pathological speakers. Note how the speakers with disarthria (top) and laryngectomy (bottom) fail to activate the alveolar and plosive PLF, respectively.}
    \label{fig:combined}
\end{figure}

While the frame-level PLFs can provide a detailed temporal view of the corresponding phonological feature as shown in Fig.~\ref{fig:combined}, an utterance-level feature representation is often necessary to describe global properties (e.g. intelligibility) of pathological speech utterances. We propose two methods to convert the $T$ frame-level PLFs $\pmb{V} \in \mathbb{R}^{F \times T}$ to an encompassing utterance-level feature, depending on the availability of a phonetic transcription.

\subsubsection{PLF Phone Error Rate}
Since our training setup of the PLFs includes phone predictions, it is possible to calculate the phone error rate~(PER) if a ground-truth phonetic transcription is available. This measurement is analogues to metrics based on the word error rate~(WER) of current ASR models, which are often employed in pathological speech analysis for intelligibility prediction. We also include the subcomponents of the PER (insertion, deletion and substitution rate) to allow for a more expressive feature. The phone posterior are derived from the PLF-to-phone conversion in path 1 depicted Fig.~\ref{fig:plf_training}.

\subsubsection{PLF Histogram}
Since the requirement of a phonetic transcription poses a limit on the applicability of the PLF phone error rate feature, we also introduce a text-independent PLF histogram feature. For each row corresponding to a PLF in $\pmb{V}$, we calculate a 20-bin normalized histogram. The mean value of the three lowest (L0 - L2) and highest bins (H2 - H0) are used as our PLF histogram features, with the addition of a middle bin (M) represented by the remaining mass:
$$
M = 1 - (L0 + L1 + L2 + H2 + H1 + H0).
$$
This creates a total of $ 21\times7=147 $ features in our text-independent PLF histogram. We expect that patients with a speech disorder will not be be able to use the full range of the PLFs, thus showing deviating outcomes from the healthy population. These deviations can subsequently be detected by downstream models for pathological speech analysis. For example, in speakers exhibiting hypernasality, the bin means of the nasal descriptor in a sufficiently long speech utterance are expected to be heavily skewed towards the H-bins.

\begin{figure}[t]
\begin{minipage}[b]{1.0\linewidth}
  \centering
  \centerline{\includegraphics[width=9cm]{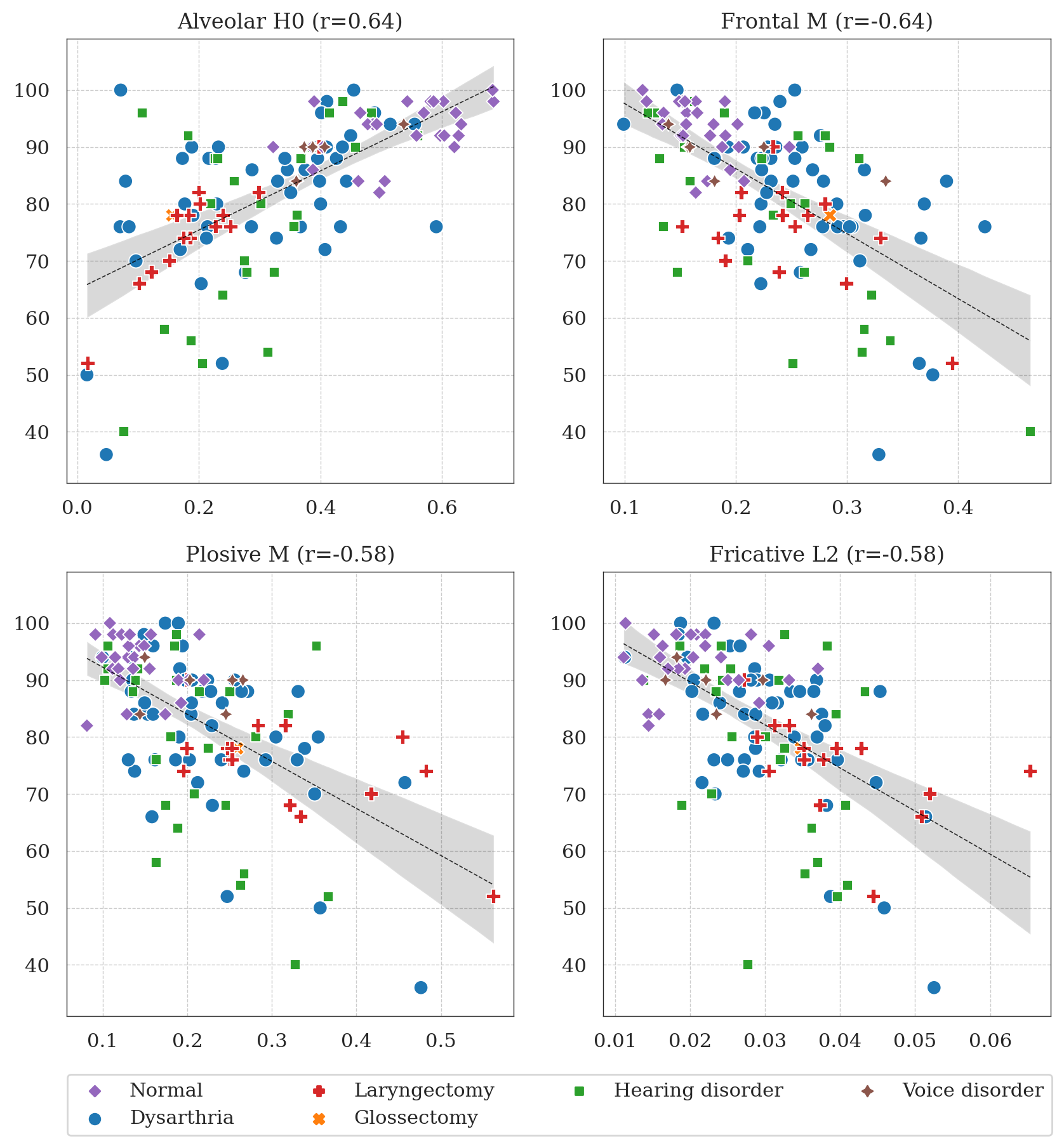}}
\end{minipage}
\caption{Correlation between best predictive PLF histogram bins and intelligibility. It also highlights the potential of PLFs in a classification task, e.g. note how the alveolar PLF groups the laryngectomy speakers.}
\label{fig:correlation}
\end{figure}

\begin{table}[h]
    \caption{Composition of speakers in COPAS dataset.}
    \label{tab:dataset}
  \centering
  \begin{tabular}{lccc}
    \toprule
    \multicolumn{1}{l}{\textbf{Category}} &
    \multicolumn{1}{c}{\textbf{Male}} &
    \multicolumn{1}{c}{\textbf{Female}} &
    \multicolumn{1}{c}{\textbf{Total}} \\
    \midrule
    Healthy & 14 & 12 & 26 \\
    \midrule
    Dysarthria & 28 & 20 & 48 \\
    Hearing disorder & 9 & 17 & 26 \\
    Laryngectomy & 15 & 0 & 15 \\
    Voice disorder & 1 & 5 & 6 \\
    Glossectomy & 1 & 0 & 1 \\
    \midrule
    \midrule
    Total & 68 & 54 & 122 \\
    \bottomrule
  \end{tabular}
\end{table}

\section{Experimental Setup}
\label{s:experimental_setup}
We test the potential of our proposed PLFs for pathological speech analysis by following the training setup described in section~\ref{ss:plf_training_setup} with the CoGeN~\cite{cogen} dataset, consisting of speech samples from 174 healthy Dutch speakers. The $E$ parameter of equation~\ref{eq:phone_prob} is set to 4 for all experiments. Subsequently, we extract the PLFs introduced in section~\ref{ss:plf_feature_extraction} from 122 Dutch speakers from the COPAS~\cite{copas} dataset, consisting of 26 normal control speakers and 96 speakers with various speech disorders. See Table~\ref{tab:dataset} for the COPAS dataset composition. The recorded speech samples consists of the speakers reading the phonetically diverse text 'Papa en Marloes' commonly used in Dutch speech therapy. In addition, an intelligibility score based on the outcome of the Dutch Intelligibility Assessment~(DIA) test~\cite{papa_marloes} is also available.



To determine the predictive capabilities of our proposed PLFs for pathological speech analysis, we train both an intelligibility prediction and speech pathology classification model using the PLFs from the COPAS dataset. In addition, we report results using other commonly used features in speech pathology analysis. We employ the \textit{emobase} feature set of the openSMILE library~\cite{opensmile}, consisting of acoustic properties and corresponding statistical derivatives of the audio signal. While being reasonably interpretable, raw acoustic properties are only of limited relevance for pathological speech analysis. Subsequently, we employ speaker embeddings from the publicly available state-of-the-art ECAPA2~\cite{ecapa2} model. Speaker embeddings are known to capture a wide variety of speaker characterizing components but its individual features are not interpretable. Finally, we use the WER of a state-of-the-art ASR system based on the publicly available XSL-R~\cite{xsl_r} architecture, fine-tuned on the Dutch subset of the Mozilla Common Voice dataset~\cite{common_voice}. ASR-based metrics provides the possibility of a linguistic analysis of the transcribed text, although in most cases this will be limited to a word-level analysis.




Due to the small size of the dataset, we employ a five-fold cross-validation setup. 20\% of the training part of each fold is used as a validation set to tune the hyperparameters of the predictive models using grid search. We treat the model type as an additional hyperparameter to ensure a fair comparison across all input features. The model types included are:  linear/logistic regression, support vector machines, decision trees and multi-layer perceptrons. Once the optimal set of hyperparameters is determined, the model is retrained using the complete training partition of the fold and evaluated on the test partition. The mean accuracy on the test partitions is reported for pathology classification with the root mean square error (RMSE) given for intelligibility prediction, with scores ranging from 0 to 100.

\begin{table}[h]
    \caption{Pathology classification (ACC) and intelligibility prediction (RMSE) performance of features on COPAS dataset.}
  \centering
  \begin{tabular}{lccccccc}
    \toprule
    \multicolumn{1}{l}{\textbf{Features}} &
    \multicolumn{1}{c}{\textbf{T-I}} &
    \multicolumn{1}{c}{\textbf{Interpretability}} &
    \multicolumn{1}{c}{\textbf{ACC}} &
    \multicolumn{1}{c}{\textbf{RMSE}} \\
    
    \midrule
     Training mean/mode & \cmark & N.A. & 0.39 & 12.88 \\
     \midrule
     openSMILE features & \cmark & medium & 0.71 & 10.01 \\
     ECAPA2 embedding & \cmark & low & \textbf{0.80} & 9.89 \\
     XSL-R WER & \xmark & medium & 0.43 & 8.78 \\
     PLF PER (\textit{ours}) & \xmark & medium & 0.59 & 8.53 \\
     PLF histogram (\textit{ours}) & \cmark & high & 0.75 & \textbf{8.43} \\
    \bottomrule
  \end{tabular}
  \label{tab:modelling_results}
\end{table}



    
    

\section{Results \& Analysis}
\label{s:results}

Table~\ref{tab:modelling_results} describes the results for pathology classification and intelligibility prediction of our proposed PLFs and other state-of-the-art features, together with a corresponding interpretability rating. For reference, the results of a weak baseline model are also given, which simply predicts the mean intelligibility score and majority pathology class of the training partitions. We observe that the baseline text-independent (T-I) features generally perform slightly worse on intelligibility prediction in comparison to our text-dependent PLF PER and XSL-R WER systems, a result often reported in other research~\cite{van_son_asr_int_cancer}. However, the pathology classification accuracy of respectively 59\% and 43\% for the PLF PER and XSL-R WER systems highlight the limited applicability of ASR transcription-based metrics for broader pathological speech analysis, given the singular nature of those features. In contrast, our proposed text-independent PLF histogram outperforms all features on intelligibility prediction and is only surpassed by the ECAPA2 embeddings on the pathology classification task. However, the PLF histograms offer superior interpretability in comparison to the latent features of the ECAPA2 embeddings, making them an overall more compelling option for pathological speech analysis.

\begin{table}[h]
  \caption{Ablation experiments of proposed PLF training setup.}
  \centering
  \begin{tabular}{lcccc}
    \toprule

    \multicolumn{1}{l}{\textbf{Experiment}} &
    \multicolumn{2}{c}{\textbf{PLF PER}} &
    \multicolumn{2}{c}{\textbf{PLF histogram}} \\

     \cmidrule(lr){2-3} \cmidrule(lr){4-5} 
    
    \multicolumn{1}{l}{\textbf{}} &
    \multicolumn{1}{c}{\textbf{ACC}} &
    \multicolumn{1}{c}{\textbf{RMSE}} &
    \multicolumn{1}{c}{\textbf{ACC}} &
    \multicolumn{1}{c}{\textbf{RMSE}} \\
    
    \midrule
Baseline  & 0.59 & 8.53 & 0.75 & 8.43 \\
\midrule
No learnable conversion matrix  & 0.53 & 9.01 & 0.70 & 9.04 \\
No direct phone classification  & 0.52 & 9.02 & 0.69 & 8.58 \\
    \bottomrule
  \end{tabular}
  \label{tab:ablation}
\end{table}

To quantify the impact of individual components of our PLF training setup, we perform an ablation study with the results given in Table~\ref{tab:ablation}. When we disable training of the learnable scalar matrix $\pmb{S}$, the intelligibility RMSE degrades with 6.4\% relative on average. This indicates that our system benefits from deviating from the strict mapping of the fixed conversion matrix $\pmb{M}$. Similarly, the 3.8\% RMSE relative degradation observed when we disable the direct phone classification path hints towards the benefits of a robust acoustic embedding to guide the learning of the PLFs.

\begin{table}[h]
  \caption{Intelligibility PCC of mean PLF value and best histogram bin.}
  \centering
  \begin{tabular}{llll|llll}
    \toprule
    
    \multicolumn{1}{l}{\textbf{PLF}} &
    \multicolumn{1}{l}{\textbf{Mean}} &
    \multicolumn{1}{l}{\textbf{Bin}} &
    \multicolumn{1}{l}{\textbf{Nr}} &
    \multicolumn{1}{l}{\textbf{PLF}} &
    \multicolumn{1}{l}{\textbf{Mean}} &
    \multicolumn{1}{l}{\textbf{Bin}} &
    \multicolumn{1}{l}{\textbf{Nr}} \\
    \midrule
Coronal & 0.50 & 0.60 & H0 & Dorsal & -0.44 & -0.57 & L1 \\
Alveolar & 0.43 & 0.64 & H0 & Nasal & -0.34 & -0.54 & L1 \\
Speech & 0.39 & 0.47 & H0 & Labial & -0.29 & 0.52 & H0 \\
Turbulent & 0.39 & 0.56 & H0 & Plosive & -0.26 & -0.58 & M \\
Mid & 0.35 & 0.42 & H0 & Diphth. & -0.20 & -0.51 & M \\
Back & 0.31 & 0.55 & H0 & Sonorant & -0.08 & -0.44 & H2 \\
Low & 0.29 & -0.37 & M & Rounded & -0.07 & 0.56 & H0 \\
Central & 0.22 & -0.50 & M & Voiced & -0.05 & 0.44 & L0 \\
Vowel & 0.03 & -0.48 & M & Lateral & -0.03 & 0.45 & H0 \\
High & 0.03 & 0.57 & H0 & Frontal & -0.03 & -0.64 & M \\
 & & & & Fricative & -0.01 & -0.58 & L2 \\
    \bottomrule
  \end{tabular}
  \label{tab:correlation}
\end{table}

To gain insight in how individual PLFs relate towards speech intelligibility, we also look at the Pearson correlation coefficient~(PCC) of the mean frame-level PLF values of each speech sample in the COPAS dataset, together with the best performing bin feature of the corresponding PLF histogram. The results are shown in Table~\ref{tab:correlation}, with the correlation of the best predictive PLFs depicted in Fig.~\ref{fig:correlation}, subdivided by speech pathology. Moderate to strong positive correlations ($ r \geq 0.4 $) with intelligibility could be found for the coronal and alveolar PLFs. Both rely on the ability of tongue displacement towards the upper teeth and roof of the mouth, possibly indicating the impact of tongue mobility on speech intelligibility. A moderately strong negative correlation ($ r = -0.44 $) was present for the dorsal PLF. Dorsal sounds rely on the back of the tongue which requires less precise movement than coronal phonemes and could be used as as compensation for the disability to produce the latter. In addition, nasality was moderately negatively ($ r = -0.34 $) associated with intelligibility which corroborates results from clinical research, where increased nasality as determined by perceptual ratings of SLTs and objective measurements using nasometers was associated with reduced intelligibility~\cite{nasal_vs_int}. In addition, we observe that the histogram-based features always show higher PCCs than their corresponding mean, indicating the importance of our proposed histogram binning strategy. While a direct comparison between our phonological features and the corresponding measurements by SLTs is missing at the moment, the current results shows the viability and additional value provided by our proposed PLFs for automatic and objective pathological speech analysis.

\section{Conclusion}
\label{s:conclusion}
We presented weakly supervised phonological features for interpretable and objective pathological speech analysis. Experiments showed that the modeling capabilities of the proposed PLFs are competitive with other state-of-the-art features while being text-independent and providing superior interpretability. In addition, our initial analysis shows promising behavior of our PLFs in comparison to the existing clinical literature regarding the relationship between intelligibility and speech characteristics.


\bibliographystyle{IEEEtran}
\bibliography{IEEEexample}

\begin{thebibliography}{10}
\providecommand{\url}[1]{#1}
\csname url@samestyle\endcsname
\providecommand{\newblock}{\relax}
\providecommand{\bibinfo}[2]{#2}
\providecommand{\BIBentrySTDinterwordspacing}{\spaceskip=0pt\relax}
\providecommand{\BIBentryALTinterwordstretchfactor}{4}
\providecommand{\BIBentryALTinterwordspacing}{\spaceskip=\fontdimen2\font plus
\BIBentryALTinterwordstretchfactor\fontdimen3\font minus \fontdimen4\font\relax}
\providecommand{\BIBforeignlanguage}[2]{{%
\expandafter\ifx\csname l@#1\endcsname\relax
\typeout{** WARNING: IEEEtran.bst: No hyphenation pattern has been}%
\typeout{** loaded for the language `#1'. Using the pattern for}%
\typeout{** the default language instead.}%
\else
\language=\csname l@#1\endcsname
\fi
#2}}
\providecommand{\BIBdecl}{\relax}
\BIBdecl

\bibitem{path_human_eval_int_non_reliable}
\BIBentryALTinterwordspacing
K.~Michi, ``Functional evaluation of cancer surgery in oral and maxillofacial region: speech function,'' \emph{International Journal of Clinical Oncology}, vol.~8, no.~1, pp. 1--17, 2003. [Online]. Available: \url{https://doi.org/10.1007/s101470300000}
\BIBentrySTDinterwordspacing

\bibitem{path_asr_2008}
M.~Windrich, A.~Maier, R.~Kohler, E.~Noeth, E.~Nkenke, U.~Eysholdt, and M.~Schuster, ``Automatic quantification of speech intelligibility of adults with oral squamous cell carcinoma,'' \emph{Folia phoniatrica et logopaedica : official organ of the International Association of Logopedics and Phoniatrics (IALP)}, vol.~60, pp. 151--6, 03 2008.

\bibitem{asr_int_overview}
\BIBentryALTinterwordspacing
M.~Karbasi and D.~Kolossa, ``Asr-based speech intelligibility prediction: A review,'' \emph{Hearing Research}, vol. 426, p. 108606, 2022. [Online]. Available: \url{https://www.sciencedirect.com/science/article/pii/S0378595522001745}
\BIBentrySTDinterwordspacing

\bibitem{asr_int_clinical_1}
F.~Stelzle, C.~Knipfer, M.~Schuster, T.~Bocklet, E.~N{\"o}th, W.~Adler, L.~Schempf, P.~Vieler, M.~Riemann, F.~W. Neukam, and E.~Nkenke, ``Factors influencing relative speech intelligibility in patients with oral squamous cell carcinoma: a prospective study using automatic, computer-based speech analysis,'' \emph{International journal of oral and maxillofacial surgery}, vol.~42, no.~11, pp. 1377--1384, 2013.

\bibitem{van_son_asr_int_cancer}
\BIBentryALTinterwordspacing
B.~M. Halpern, S.~Feng, R.~{van Son}, M.~{van den Brekel}, and O.~Scharenborg, ``Automatic evaluation of spontaneous oral cancer speech using ratings from naive listeners,'' \emph{Speech Communication}, vol. 149, pp. 84--97, 2023. [Online]. Available: \url{https://www.sciencedirect.com/science/article/pii/S016763932300047X}
\BIBentrySTDinterwordspacing

\bibitem{path_xvec_int}
S.~Quintas, J.~Mauclair, V.~Woisard, and J.~Pinquier, ``Automatic prediction of speech intelligibility based on x-vectors in the context of head and neck cancer,'' in \emph{Interspeech 2023}, 10 2020.

\bibitem{x_vectors}
D.~Snyder, D.~Garcia-Romero, G.~Sell, D.~Povey, and S.~Khudanpur, ``X-vectors: Robust dnn embeddings for speaker recognition,'' in \emph{ICASSP 2018}, 2018, pp. 5329--5333.

\bibitem{ecapa_tdnn}
B.~Desplanques, J.~Thienpondt, and K.~Demuynck, ``{ECAPA-TDNN}: Emphasized channel attention, propagation and aggregation in {TDNN} based speaker verification,'' in \emph{Proc. {INTERSPEECH} 2020 -- 21\textsuperscript{st} Annual Conference of the International Speech Communication Association}, 2020, pp. 3830--3834.

\bibitem{plfs_asr_1994}
\BIBentryALTinterwordspacing
L.~Deng and D.~X. Sun, ``{A statistical approach to automatic speech recognition using the atomic speech units constructed from overlapping articulatory features},'' \emph{The Journal of the Acoustical Society of America}, vol.~95, no.~5, pp. 2702--2719, 05 1994. [Online]. Available: \url{https://doi.org/10.1121/1.409839}
\BIBentrySTDinterwordspacing

\bibitem{jp_asr_plfs}
F.~Stouten and j.-p. Martens, ``Speech recognition with phonological features: Some issues to attend,'' in \emph{Interspeech 2006}, vol.~1, 09 2006.

\bibitem{jp_plf_asr_dependent}
C.~Middag, G.~Nuffelen, j.-p. Martens, and M.~Bodt, ``Objective intelligibility assessment of pathological speakers,'' in \emph{Interspeech 2008.}, 09 2008, pp. 1745--1748.

\bibitem{jp_plf_asr_free}
C.~Middag, Y.~Saeys, and j.-p. Martens, ``Towards an asr-free objective analysis of pathological speech,'' in \emph{Interspeech 2010}, 09 2010, pp. 294--297.

\bibitem{specaugment}
D.~S. Park, W.~Chan, Y.~Zhang, C.-C. Chiu, B.~Zoph, E.~D. Cubuk, and Q.~V. Le, ``{SpecAugment}: A simple data augmentation method for automatic speech recognition,'' in \emph{Proc. Interspeech 2019}, 2019, pp. 2613--2617.

\bibitem{cogen}
K.~Demuynck, D.~V. Compernolle, C.~V. Hove, and J.~P. Martens, ``Een corpus gesproken nederlands voor spraaktechnologisch onderzoek. final report of cogen project,'' ELIS UGent, Gent, Tech. Rep., 1997.

\bibitem{copas}
G.~V. Nuffelen, M.~D. Bodt, C.~Middag, and J.-P. Martens, ``Dutch corpus of pathological and normal speech (copas),'' \url{https://taalmaterialen.ivdnt.org/download/tstc-corpus-pathologische-en-normale-spraak-copas/}.

\bibitem{papa_marloes}
H.~Martens, G.~V. Nuffelen, and M.~D. Bodt, ``De ontwikkeling van een fonetischgebalanceerde standaardtekst,'' in \emph{NSVO-Z: Nederlands Spraakverstaanbaarheidsonderzoek - Zinsniveau}, G.~V. Nuffelen, H.~Martens, and M.~D. Bodt, Eds.\hskip 1em plus 0.5em minus 0.4em\relax Universitair Ziekenhuis Antwerpen, 2010, pp. 31--36.

\bibitem{opensmile}
F.~Eyben, M.~Wöllmer, and B.~Schuller, ``{openSMILE - The Munich Versatile and Fast Open-Source Audio Feature Extractor},'' in \emph{Proceedings of the ACM Multimedia (MM)}.\hskip 1em plus 0.5em minus 0.4em\relax Florence, Italy: ACM, October 2010, pp. 1459--1462.

\bibitem{ecapa2}
J.~Thienpondt and K.~Demuynck, ``Ecapa2: A hybrid neural network architecture and training strategy for robust speaker embeddings,'' in \emph{2023 IEEE Automatic Speech Recognition and Understanding Workshop (ASRU)}, 2023, pp. 1--8.

\bibitem{xsl_r}
A.~Babu, C.~Wang, A.~Tjandra, K.~Lakhotia, Q.~Xu, N.~Goyal, K.~Singh, P.~von Platen, Y.~Saraf, J.~Pino, A.~Baevski, A.~Conneau, and M.~Auli, ``Xls-r: Self-supervised cross-lingual speech representation learning at scale,'' \emph{arXiv}, vol. abs/2111.09296, 2021.

\bibitem{common_voice}
\BIBentryALTinterwordspacing
R.~Ardila, M.~Branson, K.~Davis, M.~Kohler, J.~Meyer, M.~Henretty, R.~Morais, L.~Saunders, F.~Tyers, and G.~Weber, ``Common voice: A massively-multilingual speech corpus,'' in \emph{Proceedings of the Twelfth Language Resources and Evaluation Conference}, N.~Calzolari, F.~B{\'e}chet, P.~Blache, K.~Choukri, C.~Cieri, T.~Declerck, S.~Goggi, H.~Isahara, B.~Maegaard, J.~Mariani, H.~Mazo, A.~Moreno, J.~Odijk, and S.~Piperidis, Eds.\hskip 1em plus 0.5em minus 0.4em\relax Marseille, France: European Language Resources Association, May 2020, pp. 4218--4222. [Online]. Available: \url{https://aclanthology.org/2020.lrec-1.520}
\BIBentrySTDinterwordspacing

\bibitem{nasal_vs_int}
K.~M. Van~Lierde, M.~De~Bodt, J.~Van~Borsel, F.~L. Wuyts, and P.~Van~Cauwenberge, ``Effect of cleft type on overall speech intelligibility and resonance,'' \emph{Folia Phoniatr Logop}, vol.~54, no.~3, pp. 158--168, May--Jun 2002.

\end{thebibliography}

\end{document}